\documentclass[aps,prl,twocolumn,showpacs]{revtex4-1}
\usepackage{amsmath}
\usepackage{amsfonts}
\usepackage{amssymb}
\usepackage{graphicx}
\usepackage{color}	

\begin{document}

\title{Effects of lasing in a one-dimensional quantum metamaterial}

\author{Hidehiro Asai$^{1,2}$}
\author{S. Savel'ev$^{2,3}$}
\author{S. Kawabata$^{1}$}
\author{A. M. Zagoskin$^{2,3}$}
\affiliation{$^{1}$Electronics and Photonics Research Institute
(ESPRIT),  National Institute of Advanced Industrial Science and
Technology (AIST), Tsukuba, Ibaraki 305-8568, Japan}
\affiliation{$^{2}$Department of Physics, Loughborough University, Loughborough, LE11 3TU, United Kingdom}
\affiliation{$^{3}$Center for Emergent Matter Science (CEMS), RIKEN, Wako-shi, Saitama 351-0198, Japan}

\begin{abstract}
Electromagnetic pulse propagation in a quantum metamaterial - artificial, globally quantum coherent optical medium - is numerically simulated. We show that for the quantum metamaterials based on superconducting quantum bits, initialized in an easily reachable factorized state,  lasing in microwave range is triggered, accompanied by the chaotization of qubit states and generation of higher harmonics. These effects may provide a tool for characterization and optimization of quantum metamaterial prototypes.
\end{abstract}

\maketitle

\section{Introduction}

The rapid development of quantum technologies since 2000 resulted in routine fabrication of solid state-based artificial quantum structures, such as qubit arrays, quantum annealers etc.\cite{Xiang2013} The achieved levels of control and global quantum coherence of these systems still fall short of the requirements of universal quantum computing, and better theoretical methods of their simulation and assessment are required\cite{Zagoskin2013a}, but they are already adequate for more the realization of such structures as, e.g., quantum metamaterials \cite{Rakhmanov2008,Quach2011,Felbacq2012,Zagoskin2011,Zagoskin2012}. These are artificial media, which   (i) are comprised of quantum coherent unit elements with desired (engineered) parameters; (ii) allow at least limited direct control of quantum states of these elements; and (iii) can maintain global coherence for the duration of time, exceeding the traversal time of an electromagnetic signal. Of course, they must also satisfy the standard requirement that the size of a unit cell of the system be much less - in practice at least twice less - than the wavelength of the relevant electromagnetic signal, so that they can be treated as a "medium". The totality of (i)-(iii) makes a quantum metamaterial a qualitatively different system, with a number of unusual properties and applications. In particular, bifocal superlens \cite{Quach2011} and quantum phase-sensitive antennas \cite{Zagoskin2013} were predicted.

Superconducting technology provides arguably the best scalability for multi-qubit systems\cite{Xiang2013,You2011}, and much of attention is therefore concentrated on superconductor-based quantum metamaterials. Experimental demonstration of a wide range of quantum-optical effects in a single superconducting flux qubit interacting with the electromagnetic field in a one-dimensional waveguide\cite{Astafiev2010,Astafiev2010a,Abdumalikov2010}, in a quantitative agreement with theoretical models, confirmed that fabrication and investigation of superconducting quantum metamaterials is well within the scope of the existing experimental techniques. The first one-dimensional superconducting quantum metamaterial prototype was recently realized\cite{Macha2013}. 

If the system is initialized in one of its macroscopic excited states, one can expect lasing. Lasing in a single artificial atom (superconducting charge qubit) was observed \cite{Astafiev2007}, and considered theoretically  \cite{Ashhab2009,Andre2009}.
Here we consider the effect in an extended system 
composed of a large number of qubits densely aligned in superconducting wave guides.
%
In this quantum metamaterial system, the lasing starts associated with correlated dynamics of qubit states due to the qubit-qubit coupling via electromagnetic field.
 We will show that in addition to the signal amplification, there appear two other peculiar phenomena: higher harmonics of the electromagnetic wave and chaotization of qubit dynamics.

\begin{figure}%
\includegraphics[width=7.5cm]{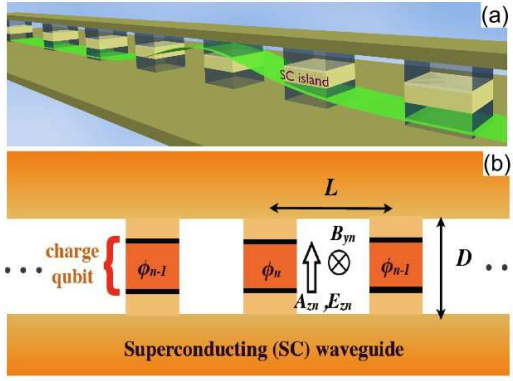}%
\caption{(a) One-dimensional (1D) quantum metamaterial based on charge qubits placed in a waveguide formed by two bulk superconductors.
 The charge qubit is formed by a superconducting island, separated from the bulk superconductors by identical Josephson junctions,
 and play the role of artificial atoms. The green wave represents distribution of the electromagnetic wave in the metamaterial.
 The control circuitry of qubits (gate electrodes and bias current sources) is not shown.
 (b) The cross-sectional view of the 1D quantum metamaterial.  The superconducting waveguides are separated with the distance $D$. The charge qubits align at regular intervals $L$. $\phi_n$ is the superconducting phase of the $n$th qubit island. 
 $A_{z,n}$,  $B_{y,n}$ and $E_{z,n}$ are the vector potential, the magnetic field and the electric field of the $n$th unit cell, respectively.
}%
\label{fig1}%
\end{figure}

Due to the combination of large qubit dipole moment and small field-mode volume, one-dimensional superconducting metamaterials are characterized by an exceptionally strong coupling between the "atom" (qubit) and the electromagnetic field mode. The ratio of coupling strength to the mode frequency, $g/\omega \sim 0.02$, while in the case of a real (optical range) or artificial (microwave range) atom coupled to a three-dimensional field mode it is of order $10^{-7}$  (see \cite{Zagoskin2011}, Table 4.1 (p.189)). This increases the range of effects, which can be observed in the system, leading to the effective suppression of decoherence. On the other hand, the usefulness of the perturbation theory, employed in the early research of the subject \cite{Rakhmanov2008,Zagoskin2008}, is limited to perturbative dynamics expected in weak coupling condition. Therefore, in one-dimensional quantum metamaterials of this kind, we can neglect losses and decoherence, but must rely on numerical solution of the equations describing the pulse propagation in the system.

As in \cite{Rakhmanov2008, Shvetsov2013}, we will treat the electromagnetic field classically, and approximate the quantum state of the quantum metamaterial by a product of single-qubit wave functions. The latter approximation requires some justification, because using it we lose all the effects of entanglement between qubits and drastically reduce the complexity of states available to the system. Indeed, such an approximation does not allow to consider, e.g., quantum birefringence, i.e., the system being in a superposition of states $|000...0\rangle$ and $|111...1\rangle$ (which correspond to different refractive indices), but it is adequate for the description of lasing. This is fortunate, since such factorized states are the most robust of quantum coherent states of a macroscopic system, being, in a sense,  the closest to classical ones \cite{Leggett1999,Leggett1999a}. Their decoherence time is limited by the decoherence time of a single qubit, and can therefore greatly exceed the period of field oscillation.  In the continuum limit \cite{Rakhmanov2008,Zagoskin2011} such states are described by a two-component macroscopic wave function $\hat{\Psi}(x)$, formally analogous to superconducting order parameter $\Delta(x)$. The crucial distinction between them comes from the fact that  the former is transitional and eventually disappears, while the latter is an equilibrium quantity, nonzero value of which is supported and protected from destruction by thermal fluctuations by the interactions in the system. 

\section{Model}
We do not expect that the character of signal propagation in a quantum metamaterial will qualitatively depend on whether it is built of charge or flux qubits \cite{Zagoskin2011} (or, for that, any other kind of superconducting qubit).
 We will therefore consider here a quantum metamaterial based on charge qubit as the unit element \cite{Rakhmanov2008,Shvetsov2013} (Fig.~\ref{fig1}) (the proof-of-principle lasing experiment \cite{Astafiev2007} used a single charge qubit). The separation $D$ between the superconducting waveguides and the period $L$ of the structure are much less than the wavelength of the electromagnetic field mode in the waveguide. This allows us to neglect the variation of electric field and vector potential within each cell and express the (classical) magnetic and electric field in the  $n$th cell as 
\begin{eqnarray}
B_{y,n} &=& \frac{A_{z,n+1}-A_{z,n}}{L};\label{eq:1} \\
E_{z,n} &=& \frac{1}{2} \frac{V_{z,n}^u - V_{z,n}^l}{D}  \nonumber \\
&=& \frac{1}{2D} \Bigl[\frac{\hbar}{2e}\frac{d}{dt}\left(0-\phi_n-\frac{2\pi}{\Phi_0}\int A_{z,n}dz\right)  \nonumber \\
&+& \frac{\hbar}{2e}\frac{d}{dt}\left(\phi_n-0-\frac{2\pi}{\Phi_0}\int A_{z,n}dz\right)  \Bigr]  \nonumber \\
&=& -\frac{h}{2e\Phi_0} \frac{d A_{z,n}}{dt},
\label{eq:2}
\end{eqnarray}
where $\phi_n$ is the superconducting phase of the $n$th qubit island, and $V_{z,n,u}$ and $V_{z,n,l}$ are the voltages on upper (lower) Josephson junctions of this qubit. The energy of the system can thus be written as the sum over unit cells (we assume that all qubits are identical), ${\cal E}_{tot} = \sum_n {\cal E}_n$, where
\begin{eqnarray}
{\cal E}_n &=& \frac{\hbar^2 C}{8e^2} \biggl[\bigl(\frac{d\phi_n}{dt}+\frac{\pi D}{\Phi_0} \frac{dA_{z,n}}{dt}\bigr)^2 + \bigl(\frac{d\phi_n}{dt}-\frac{\pi D}{\Phi_0} \frac{dA_{z,n}}{dt}\bigr)^2\biggr]  \nonumber\\
&-&  E_J\left[\cos\left(\phi_n+\frac{\pi D A_{z,n}}{\Phi_0}\right) + \cos\left(\phi_n-\frac{\pi D A_{z,n}}{\Phi_0}\right)\right]  \nonumber\\
&+& \frac{DL}{8\pi} \left[\frac{A_{z,n+1}-A_{z,n}}{L}\right]^2 + 2 I_n\phi_n \nonumber\\
&=& \frac{E_J}{2\omega_J^2}\left[\left(\frac{d\phi_n}{dt}\right)^2 + \left(\frac{\pi D}{\Phi_0} \frac{dA_{z,n}}{dt}\right)^2\right] \nonumber\\
&-& 2E_J \cos\phi_n \;\cos\frac{\pi D A_{z,n}}{\Phi_0} +  \frac{DL}{8\pi} \left[\frac{A_{z,n+1}-A_{z,n}}{L}\right]^2    \nonumber\\
&+& 2 I_n\phi_n 
\label{eq:5}
\end{eqnarray}
In Eq.~(\ref{eq:5}) the first term is the electrostatic energy, the second the Josephson energy, the next is the magnetic field energy, and the last one describes the effect of a bias current through the qubit island (as one method of direct control of its state). Further, $E_J = I_c\Phi_0/2\pi c$ is the Josephson energy; $\omega_J^2 = 2eI_c/\hbar C$ is the Josephson plasma frequency; $I_c$ and $C$ are the critical current and the capacitance of the Josephson junctions. The natural length unit is now $\lambda = c/\omega_J$. In the following we will use the dimensionless quantities $l = L/\lambda$ (unit cell length), $a_{z,n}=\pi D A_{z,n}/\Phi_0$ (vector potential), $e_{z,n} = \pi D\lambda E_{z,n}/\Phi_0$ (electric field),  $b_{y,n} = \pi D\lambda B_{y,n}/\Phi_0$ (magnetic field), $\tau = \omega_J t$ (time), $\gamma_n = I_n/I_c$ (current), and $E_n = {\cal E}_n/E_J$ (energy). This allows us to rewrite (\ref{eq:5}) as

\begin{eqnarray}
E_{tot} =\sum_n E_n &\equiv& \sum_n \left\{  E^{\textrm{qb}}_n(\phi_n,a_{z,n}) + E^{\textrm{field}}_n(a_{z,n})  \right\}  \\
E^{\textrm{qb}}_n(\phi_n,a_{z,n}) &=& \left( \frac{d\phi_n}{d\tau} \right)^2 - 2 \cos\phi_n \cos\!a_{z,n} - 2 \gamma_n\phi_n \nonumber \\
E^{\textrm{field}}_n(a_{z,n}) &=&  \beta^2\left(\frac{a_{z,n+1}-a_{z,n}}{l}\right)^2 + \left(\frac{d a_{z,n}}{d\tau}\right)^2 \nonumber
\label{eq:6}
\end{eqnarray}
where the dimensionless parameter
\begin{equation}
\beta = \frac{L\Phi_0^2}{8\pi^3D\lambda^2E_J}
\label{eq:6a}
\end{equation}
characterizes the speed of light in the metamaterial.

The  quantization of (4) is straightforward: the qubit energy $E^{\textrm{qb}}_n$ is replaced by the qubit Hamiltonian ${\cal H}_n^{\textrm{qb}}$. In order to do so, we note that the time derivative of the qubit phase in (4) is related to the charge on the island via
$$ Q_n = -2eN_n = CV_{z,n}^u - CV_{z,n}^l = -\frac{\hbar C}{e} \frac{d\phi_n}{dt},$$
so that the quantization is achieved by the substitution \cite{Zagoskin2011}
\begin{equation}
\frac{d\phi_n}{dt} = \frac{2e^2}{\hbar C}N_n \to \frac{2e^2}{\hbar C}\hat{N}_n = \frac{2e^2}{\hbar C} \frac{1}{i}\frac{\partial}{\partial \phi_n}.
\label{eq:8}
\end{equation}
(The action of a gate potential $V_g$ - another method of controlling the qubit state - can be taken into account by replacing the right-hand side of (\ref{eq:8}) with 
$\frac{2e^2}{\hbar C}(\hat{N}_n-n^*),$ where $n^* = C_gV_g/2e$.) Now the $n$th qubit's Hamiltonian takes the form
\begin{eqnarray}
%
{\cal H}_n^{\textrm{qb}} (\phi_n, a_{z,n}) &=& -  \frac{2 e^2}{C} \frac{\partial^2}{\partial\phi_n^2} - 2E_J \cos\phi_n \cos a_{z,n} - 2\gamma_n\phi_n \nonumber\\
 &=& {\cal H}_{n,0}^{\textrm{qb}}(\phi_n) + {\cal H}_{n,\textrm{int}}^{\textrm{qb}}(\phi_n, a_{z,n}),
\label{eq:9}
\end{eqnarray}
where we have split it in the unperturbed part,
\begin{equation}
{\cal H}_{n,0}^{\textrm{qb}} (\phi_n)= - \frac{2 e^2}{C} \frac{\partial^2}{\partial\phi_n^2} - 2E_J \cos\phi_n - 2\gamma_n\phi_n,
\label{eq:10a}
\end{equation}
and the field-dependent perturbation,
\begin{equation}
{\cal H}_{n,\textrm{int}}^{\textrm{qb}}(\phi_n, a_{z,n}) = 2E_J \cos\phi_n (1-\cos a_{z,n}).
\label{eq:10}
\end{equation}

For the above mentioned reasons, we limit our considerations to the factorized quantum states of the system, 
\begin{equation}
|\Psi \rangle = \bigotimes_n |\Psi_n \rangle = \bigotimes_n \left(C_0^n |0\rangle + C_1^n |1\rangle\right)
\label{eq:10b}
\end{equation}
(we do not explicitly label by $n$ the states of the $n$th qubit, $|0\rangle$ and $|1\rangle$, since they always appear accompanied by the $C^n$'s). 
Therefore we can solve for each unit cell separately, regarding the vector potential as a classical parameter. Since the states   $|0\rangle$ and $|1\rangle$ of a charge qubit differ by one Cooper pair, the only nonzero matrix elements of the operator $\cos\phi_n$ in the basis $\{|0\rangle, |1\rangle\}$ are $\langle 0|\cos\phi_n|1\rangle = \langle 1|\cos\phi_n|0\rangle = 1$ \cite{Zagoskin2011}. Therefore for the coefficients $C_{0,1}^n$ in the Schr\"{o}dinger representation we find:
\begin{eqnarray}
 \label{eq:11}
i\frac{d}{d\tau} C_0^n &=& \varsigma (1-\cos a_{z,n}) C_1^n \exp[-i\varsigma\epsilon\tau];\nonumber\\
i\frac{d}{d\tau} C_1^n &=& \varsigma (1-\cos a_{z,n}) C_0^n \exp[i\varsigma\epsilon\tau]. 
\end{eqnarray}
Here $\epsilon = ({\cal E}_1-{\cal E}_0)/E_J$ is the dimensionless qubit excitation energy, and $\varsigma = E_J/\hbar\omega_J$.

Now, following the general approach of \cite{Rakhmanov2008,Zagoskin2011}, we write the Hamilton function $\langle H\rangle$ for the classical variable $a_{z,n}$ by
replacing the qubit energy $E_n^{qb}$ in the expression (4) for the total energy of the system with their quantum expectation values, i.e., with the matrix elements,
\begin{equation}
E_n^{\textrm{qb}} \  \rightarrow  \ \langle\Psi |\hat{H}_n^{qb}(\phi_n,a_{z,n})|\Psi \rangle
\label{eq:12}
\end{equation}   
(here $\hat{H} = {\cal H}/E_J$). The canonical momentum $\Pi_n = 2 \frac{\partial a_{z,n}}{\partial\tau} $. The vector potential then satisfies the Hamilton equations
\begin{eqnarray}
& &\frac{\partial}{\partial\tau} \Pi_{n}= 2 \frac{\partial^2  a_{z,n}}{\partial\tau^2} = - \frac{\partial \langle H\rangle}{\partial a_{z,n}}  \nonumber\\
&=& -2V_n \sin a_{z,n} + 2\beta^2 \frac{a_{z,n+1}-2a_{z,n}+a_{z,n-1}}{l^2}.
\label{eq:13}
\end{eqnarray}
Here 
\begin{equation}
V_n = C_0^{n*}C_1^{n}e^{-i \varsigma\epsilon\tau} + C_0^{n}C_1^{n*}e^{i \varsigma\epsilon\tau}.
\label{eq:14}
\end{equation}

Within the assumptions we made, equations (\ref{eq:11}) and (\ref{eq:13}) fully describe the evolution of the system. In the continuous limit, $l \to 0$, they would reduce to the wave equation for the vector potential coupled to the two-component field $\Psi_{0,1}(x,\tau)$,  supplemented by the equations for $\Psi_{0,1}(x,\tau)$. Using realistic parameters of the system, one can see that the dissipation is negligible on the time scale of the effects we are interested in (see \cite{Rakhmanov2008}), and are therefore neglected. 
If we take $E_J = 10 \sim 50 \mu eV$ and $E_c = 10 E_J$~\cite{Nakamura1999},
$\omega_J$ becomes $\sim 10^{11}$ (1/s). In this case, $\tau= 10^4 $ corresponds to the typical decoherence time for charge qubits 100 ns, and 
our lasing occurs $\tau \sim 10^3 \ll 10^4 $. Thus, we believe our simulation is justified.

For a better efficiency of numerics, Eq.~(\ref{eq:11}) are split into real and imaginary parts and solved using the 4th order Runge-Kutta method.  Eq.~(\ref{eq:13}) for the vector potential is rewritten in terms of dimensionless electric and magnetic fields as
\begin{eqnarray}
\frac{de_{z,n}}{d\tau} = \beta^2\frac{b_{y,n}-b_{y,n-1}}{l} + V_n \sin a_{z,n}; \nonumber\\
b_{y,n} = -\frac{a_{z,n+1}-a_{z,n}}{l}; \:\: e_{z,n} = -\frac{da_{z,n}}{d\tau}; \\
\frac{db_{y,n}}{d\tau} = -\frac{e_{z,n+1}-e_{z,n}}{l}, \nonumber  
\label{eq:15}
\end{eqnarray}
and the resulting system is solved using finite difference time domain method.

\begin{figure}%
\includegraphics[width=7.5cm]{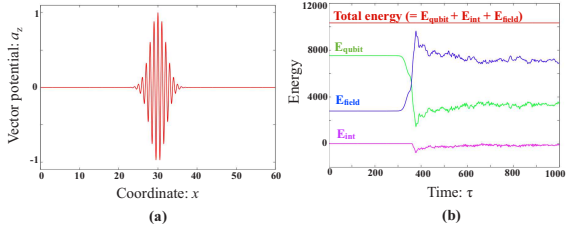}%
\caption{(a) The initial distribution of vectorpotential $a_z$ in a unit cell.(Gaussian pulse)
 (b) Time evolution of energy in the system. Bottom to top:
 field-qubit interaction energy $E_{\textrm{int}}$ (mauve online);
 electromagnetic field energy $E_{\textrm{field}}$ (blue online);
 qubit energy $E_{\textrm{qubit}}$(green online);
total energy $E_{\textrm{total}}$(red online).  }%
\label{fig:2}%
\end{figure}

\section{Results and conclusions}

We consider a quantum metamaterial containing $N = 1200$ unit cells, with periodic boundary conditions: 
\begin{equation}
a_{z,0} = a_{z,N-1};\: a_{z,1} = a_{z,N}.
\label{eq:16}
\end{equation}
We chose the following values dimensionless parameters for the metamaterial:  $\varsigma = 1$, $\epsilon = 2\pi$, $\beta = 1$, $l = 0.05$. Its initial state is fully excited: $C_0^n(0) = 0; \: C_1^n(0) = 1.$ 

The initial state of the electromagnetic field is a Gaussian pulse,
\begin{equation}
a_{z,n}(0) = P \exp\left[-\frac{(x_n+\beta\tau)^2}{2q^2}\right] \cos(k_g x_n-\omega_g\tau),
\label{eq:17}
\end{equation}
where $\omega_g = k_g\beta$, and we choose $P = 1, q=2$, and $\omega_g = \epsilon = 2\pi$. 
$x_n = nl$ is the coordinate of the $n$th qubit along the $x$ direction.
Fig.~\ref{fig:2} (a) shows the initial distribution of $a_{z,n}(0)$ in a unit cell ($x=0\sim60$).
With the chosen parameters, the wave packet traverses the system within $\tau_0 = 60$.
 Our use of cyclic boundary conditions allows us to investigate the system's behaviour on longer times, in the regime analogous to generation regime in a laser with a positive feedback loop.

\begin{figure}%
\includegraphics[width=7.5cm]{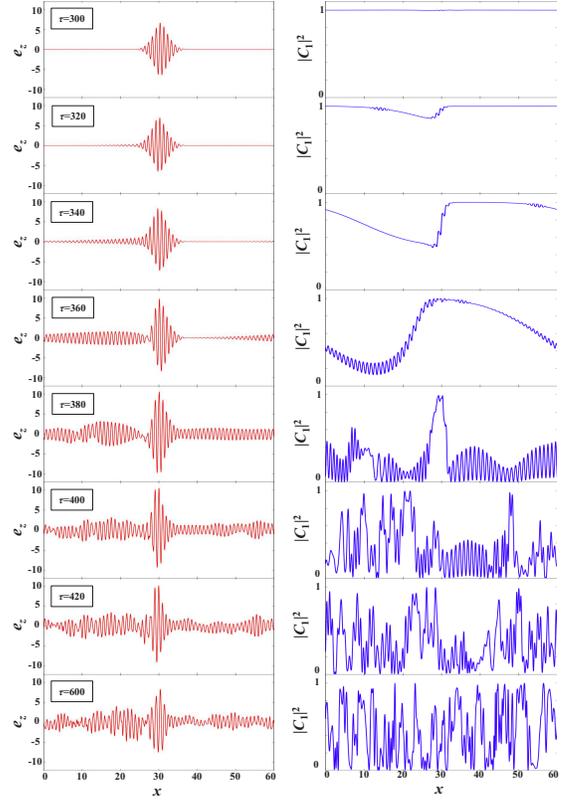}%
\caption{Electric field amplitude (left panel) and the probability amplitude of the excited state (right panel) in the metamaterial at $\tau = 300; 320; 340; 360; 380; 400; 420$ and $600$ (top to bottom). The origin of $x$ is shifted to keep the wave packet at the centre of the panel.}%
\label{fig:3}%
\end{figure}

First we reassure ourselves that lasing actually occurs from 
the change of the component of the system energy.    
Figure \ref{fig:2} (b) shows the time evolution of respective energies; 
the qubit energy $E_{\textrm{qubit}} = \sum_n \langle \Psi_n| H_{n,0}^{\textrm{qb}} |\Psi_n \rangle$,
the field-qubit interaction energy $ E_{\textrm{int}}= \sum_n \langle \Psi_n| H_{n,\textrm{int}}^{\textrm{qb}} |\Psi_n \rangle$,  
electromagnetic field energy $ E_{\textrm{field}}= \sum_n E^{\textrm{field}}_n $,
and the total energy $E_{\textrm{total}} =E_{\textrm{qubit}} + E_{\textrm{int}} + E_{\textrm{field}}$.

In Fig.~\ref{fig:2} (b), we see that indeed at $\tau \sim 300$ the energy of excited qubits starts pumping energy into the field, and eventually the field and the qubits almost exactly exchange their energies. The total energy of the system, of course, remains constant. When looking at the electric field distribution and the quantum state of qubits as a function of $\tau$ (Fig.~\ref{fig:3}), we see that at $\tau = 320$ the field absorbs some of the energy of qubits in its vicinity. 
As can be seen from these figures, the correlated dynamics of qubit states occurs immediately after the lasing starts ($\tau = 300 \sim 360$) . 
This more or less ordered behavior of the quantum metamaterial wave function after several rounds is replaced by a chaotic dependence $|C^n_1|$ on the coordinate. By that time the energies of field and qubits stabilise. 

\begin{figure}%
\includegraphics[width=7.5cm]{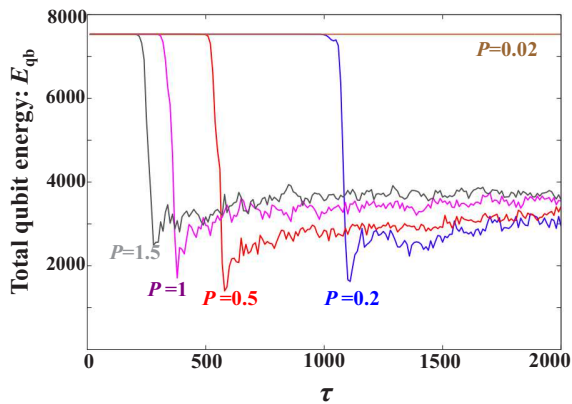}%
\caption{Qubit energy $E_{\textrm{qubit}}$ as a function of time for various field amplitudes. Right to left: $P = 0.05$ (green online); 0.2 (blue); 0.5 (pink); 1 (turquose); 1.5 (grey). At $P = 0.02$ (brown) lasing did not start up to $\tau = 40000$.}%
\label{fig:4}%
\end{figure}

\begin{figure}%
\includegraphics[width=7.5cm]{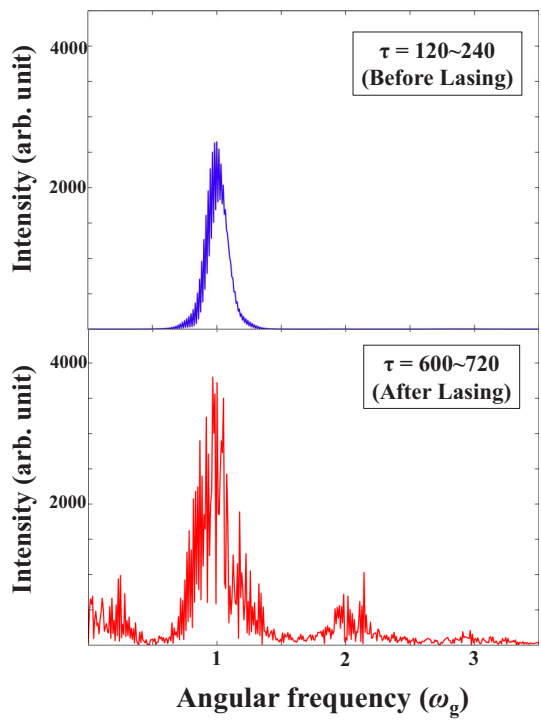}%
\caption{Field frequency spectrum for $\tau=120 \sim 240$ (top) and for $\tau=600 \sim 720$ (bottom). Note the appearance of both higher harmonic (at $\sim 4\pi = 2\omega_g$) and subharmonic peaks after the lasing occurred. }%
\label{fig:5}%
\end{figure}

The numerical calculations established one more interesting feature: the moment when lasing starts depends on the initial field amplitude (see Fig. 4).
In Fig. 4, which shows time evolution of the quibit energy $E_{\textrm{qubit}}$ for various field amplitude ($P = 0.02 \sim 1.5$),
 we see that the the moment of lasing becomes early as the amplitude increases. Provisionally, we attribute this to the rigidity of the qubit system, produced by the effective qubit-qubit interaction through the field mode.
A stronger qubit-qubit coupling should lead to their more correlated behavior and increase the amplitude of stimulated emission from the quantum metamaterial.

\begin{figure}%
\includegraphics[width=7.5cm]{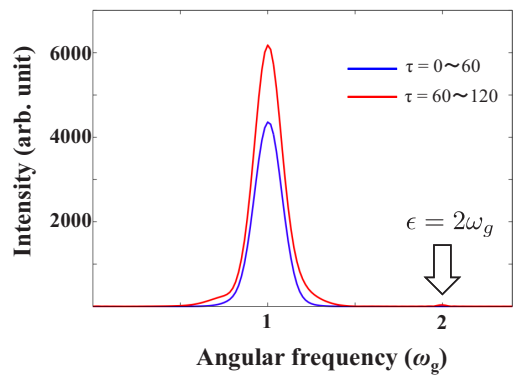}%
\caption{ Field frequency spectrum for$\tau=0 \sim 60$ (red online) and for $\tau=60 \sim 120$ (blue online) in the case of $\epsilon = 2\omega_g = 4\pi$.
 The sub-harmonic component $\frac{\epsilon}{2} = \omega_g $ increases due to parametric lasing.}%
\label{fig:6}%
\end{figure}

So far we considered the exact resonance, $\epsilon = \omega_g$. The appearance of signal at a higher harmonic and subharmonic (see Fig.~\ref{fig:5} for frequency spectrum of the electric field before and after lasing.) indicates that lasing should occur as well at other relations between $\epsilon$ and $\omega_g$. Indeed, e.g. at $\epsilon = 2\omega_g = 4\pi$  we see a clear case of parametric lasing; in this particular case it is driven by two-photon processes. Figure ~\ref{fig:6} shows the frequency spectrum of the electric field for the case of $\epsilon = 2\omega_g = 4\pi$ (Note: other calculation parameters are same as the aforementioned calculations.). In this case, lasing immediately occurs  after the pulse starts to propagate, and the component of the sub-harmonic frequency $\frac{\epsilon}{2}$ increases as shown in this figure.
Here, as before, after some transition period the qubit and field subsystems exchange their energies and 
reaches to stationary states
when the qubit states (not shown) are chaotized.

In conclusion, we have investigated the lasing effects in a one-dimensional quantum metamaterial based on superconducting  charge qubits by numerically solving the equations of motion for the system in the presence of a classical electromagnetic pulse.
 In case when  the signal can propagate repeatedly through the system,
the lasing occurs associated with a correlated behavior of qubit states,
  and the electromagnetic pulse and the qubit subsystem exchange their energies.
The onset of lasing strongly depends on the pulse amplitude, 
and this is resulting from the effective rigidity of the qubit subsystem, interacting through the electromagnetic mode.
  After several rounds of the pulse propagation, system reaches stationary regime, 
  which is characterized by a chaotic qubit state distribution.
 Nonlinear coupling of qubits to the field leads to efficient parametric lasing, e.g., when the pulse frequency is only a half of the qubit interlevel distance. 
 The system turns out to have rich physics, within the range of parameters currently available for experimental investigation. 
 Moreover, our quantum metamaterial works as an nonlinear laser medium whose parameters can be artificially controlled,
 and will be a promising candidate for practical metamaterial.

We are grateful to Dr. M. Everitt and Prof. T. Kato for many fruitful discussions. 
H.~A. and S.~K. are partially supported by by a Grant-in-Aid for JSPS Fellows, 
a Grant-in-Aid for Scientific Research from the Ministry of Education, Science, Sports and Culture of Japan (Grants No. 24510146 and No. 26790062)


\end{document}